# THE SUN IS A MAGNETIC PLASMA DIFFUSER
# THAT SORTS ATOMS AND ELEMENTS BY WEIGHT

(With web links to atomic weight measurements that unmasked the Iron Sun)

O. Manuel, Nuclear Chemistry, University of Missouri-Rolla, MO 65401 USA
http://web.umr.edu/~om/abstracts2005/The_Suns_Origin.pdf

The fusion of hydrogen atoms – the same mechanism that is at the heart of a Hydrogen Bomb – is generally accepted as the energy source of our Sun. Yet, if this theory was the actual source of the Sun's energy,

a) *Why is the Sun's emission of neutrinos only about 1/3 the amount expected from fusion reactions?*

b) *Why is the ratio of Oxygen atoms to Carbon atoms only 2 at the Sun's surface*, when laboratory and theoretical calculations predict a much higher value?

c) If fusion powers the Sun, *why does it discard 50 trillion tons of Hydrogen each year as "solar wind" trash, with traces of other elements carefully sorted by atomic weight?*

d) *How are we to understand the universe when*, as Nobel Laureate William A. Fowler noted in 1988, *". . . we do not even understand how our own star really works."* [1]?

These are a few of the questions answered in two new papers [2,3]. The papers also confirm major findings of a 45-year forensic investigation of the "fingerprints" left in the atomic weights of elements in meteorites, planets, the solar photosphere, and in material ejected by the solar wind and solar flares.

The new results affirm Theodore W. Richards' Nobel Lecture on Atomic Weights, 6 Dec 1919: *"If our inconceivably ancient Universe even had any beginning, the conditions determining that beginning must even now be engraved in the atomic weights."*

*Atomic weights are like solar DNA. They are the average weights of the atoms that comprise each element. Atomic weights provide a detailed fingerprint of the star that gave birth to the solar system and they tell us what is happening inside the Sun today.*

Here are the main conclusions to our study, with numbered references to web sites:

**I. THE SUN IS A GIANT PLASMA "DIFFUSER"** or "sorter of ionized atoms (elements and isotopes) by weight". This explains why lightweight elements - Hydrogen and Helium - cover the Sun's surface while the interior of the Sun is made of elements that are common in meteorites, Earth, and in other rocky planets close to the Sun [2, 3].

**a**. In 1983 Manuel and UMR graduate student, Golden Hwaung, reported that a total of 22 different types of atoms in the solar wind had been sorted by weight [4]. Those atoms weighed from 3 to 136 times the weight of Hydrogen, the lightest element. When the abundances of elements at the Sun's surface were corrected for this sorting, the interior of the Sun was found to consist almost entirely of seven, even-numbered elements – Iron, Oxygen, Silicon, Nickel, Sulfur, Magnesium and Calcium.

{Three tests of this hypothesis were proposed in 1983 (See [4], p. 13). They are the basis for the ill-fated *Genesis Mission* using collector foils to capture elements from the solar wind, for numerous solar neutrino detectors including the *Sudbury Neutrino Observatory*, and for mass-spectrometer measurements on the *Galileo Probe of Jupiter*.}

**b**. In 2002 Professor Stig Friberg (Clarkson University) and Manuel noted at the SOHO Helio-Seismology conference that 99% of the material in ordinary meteorites is made of these same elements. Dr. "Sam" Samaranayake, UMR Associate Professor of Statistics, helped them show [5] that the *likelihood of this spectacular agreement being a meaningless coincidence is less than one in 50,000,000,000,000,000,000,000,000,000,000*.

**c**. The new study compared the amounts of 72 different types of atoms in the solar photosphere with those predicted from nuclear reactions to confirm that ionized atoms are sorted by weight in the Sun. Those atoms weigh 25 to 207 times the weight of Hydrogen atoms. These new results [2] were submitted for presentation at the special Genesis Session of the 36[th] Lunar & Planetary Science Conference in Houston, TX.

**d**. Although Earth's gravitational field is 330,000 times weaker than the Sun's field, even here the lightest gas, Hydrogen, is sorted out and moved to the top of the atmosphere while heavier Carbon-dioxide gas is concentrated in low-lying places like Death Valley. However the sorting of ionized atoms in the Sun is caused primarily by a plasma flow of magnetically guided protons, rather than by gravitational fractionation [4].

**II. SOLAR MAGNETIC FIELDS ACCELERATE PROTONS (HYDROGEN IONS) UPWARD** and create the "carrier gas" that sorts ions in the Sun by weight [6]. Solar magnetic fields from the Sun's iron-rich interior also cause eruptions at the solar surface and intermittent changes in our climate. These are some of the conclusions to a 2003 study [6] by Professors Stig E. Friberg (Clarkson University), Barry W. Ninham (University of Florence and Australian National University) and Manuel.

**a**. Solar magnetic fields may determine the effectiveness of the sorting process. A recent survey [7] found iron-rich surfaces of stars during periods of low magnetic activity, like the Sun's "Maunder minimum" that accompanied the "Little Ice Age" in Europe.

**b**. Magnetic fields that penetrate the solar surface bring fast-moving, less-sorted, heavy ions (elements & isotopes) up from the interior of the Sun as the solar energetic particles (SEP) observed in violent solar flares and eruptions [8, 9].

**c**. Oxygen atoms are 33% heavier than Carbon atoms. The second new paper [3] shows that plasma sorting of ionized atoms (elements and isotopes) by weight causes the ratio of Oxygen to Carbon to be only 2 at the surface of the Sun, although the O/C ratio is 10 inside the Sun. This explains the other fundamental problem that Nobel Laureate

William A. Fowler identified in the field of nuclear astrophysics in 1988, *". . . we still cannot show in the laboratory and in theoretical calculations why the ratio of oxygen to carbon in the sun and similar stars is close to two-to-one"* [1].

**III**. **FRESH, RADIOACTIVE SUPERNOVA DEBRIS FORMED METEORITES, PLANETS, AND THE SUN** [5, 10, 11]. In addition to the radioactive elements (like Th, U, and K) that survive and still make the Earth's insides hot today, 5 billion years after the supernova exploded, decay products of short-lived species like iodine-129 and plutonium-244 are seen in iron meteorites and in the interior of the Earth [12].

**a**. Iron-rich material from deep within the supernova directly formed iron meteorites, the interior of the Sun, and the iron cores of planets near the Sun [5, 10, 11].

b. The Sun formed on the collapsed supernova core, a pulsar [5, 6]

**c.** Material near the surface of the supernova formed giant, gaseous planets like Jupiter out of lightweight elements like Hydrogen, Helium, Carbon and Nitrogen [10, 11]. Those elements did not mix with the elements that made the Sun. Undergraduate students used data from the *Galileo Mission* to Jupiter to confirm this. They showed that i) the atomic weight of Xenon in Jupiter is "strange", unlike Xenon in the Earth and the Sun [13], and ii) nuclear reactions in the Sun could not change Jupiter-type Hydrogen and Helium into the Hydrogen and Helium seen in the solar wind today [14].

**IV**. **THE SUN IS A "CLOTHED NEUTRON STAR"** formed by accreting iron-rich material on the pulsar (spinning neutron star) made at the supernova core [5, 6].

**a**. Students in Manuel's Advanced Nuclear Chemistry class, Chem. 471, in the spring of 2000 uncovered evidence that n-n repulsion causes neutrons to be energetically ejected, rather than "drip," from neutron-rich material. Later Manuel and co-workers showed that neutron-emission generates most of the Sun's energy [15, 16]. In this process a neutron is ejected from the Sun's central neutron star with the release of 10-22 MeV of energy:

<neutron> **-->** free neutron + 10-22 MeV

The free neutron then decays to a proton and an electron:

free neutron **-->** proton + electron + 0.782 MeV;   $p^+ + e^- = H$ (Hydrogen)

**b**. *Hydrogen is a by-product or "smoke" from the nuclear furnace at the Sun's core*. Hydrogen is mostly consumed by fusion reactions as it moves upward in the solar "flue." Hydrogen fusion generates less than 38% of the Sun's energy [15, 16].

**c**. *Solar neutrinos are not "missing"* [17]. "Smoke" was mistaken for solar fuel. That misunderstanding produced the "Solar Neutrino Puzzle", the other major problem that Nobel Laureate William A. Fowler identified in nuclear astrophysics in 1988 [1].

**d**. *The Sun operates like high efficiency furnaces with catalytic converters that burn most of the "smoke" from the nuclear furnace before it exits the flue*. Hydrogen pouring from the Sun's surface is "smoke" from the furnace that powers the Sun. Each year 50 trillion metric tons of protons (Hydrogen ions) reach the Sun's surface and are flung out into space by the solar wind. This is a small fraction of the Hydrogen "smoke" (protons + electrons) generated in the nuclear furnace at the Sun's core [18].

These findings resolve two problems Noble Laureate William A. Fowler identified in the most basic concepts of nuclear astrophysics [1]:

1. The solar neutrino problem indicates that *". . . we do not even understand how our own star really works.",* and

2. *". . . we still cannot show in the laboratory or in theoretical calculations why the ratio of Oxygen to Carbon in the Sun and similar stars is close to two to one . . ."*

Solving these mysteries of the Sun opens the door to many others. The Sun is the model for other stars in the cosmos and it accounts for > 99.8% of the mass of the solar system.

Many students and colleagues contributed to these findings. Mr. Sumeet Kamat, a graduate student in Computer Science, is putting the names and photographs of some of those who contributed to the effort on Manuel's web page: http://www.umr.edu/~om

Fate played a major role in these findings, dropping a meteorite filled with Neon from the Sun near Fayetteville, Arkansas in 1934, when Kazuo Kuroda was a 17-year old student in Tokyo and Manuel was not yet conceived. Twenty-eight years later, University of Arkansas Professor Kuroda handed the Fayetteville meteorite to his graduate student, O. Manuel, for analysis. Manuel carried the sample to UC-Berkeley and found puzzling evidence that the atomic weight of neon varied in the Fayetteville meteorite: The three types of Neon atoms in the meteorite - $^{20}$Ne, $^{21}$Ne, and $^{22}$Ne - had been sorted by weight.

Almost a half-century after the meteorite landed near Fayetteville, and dozens of analyses showing that Neon atoms in many meteorites had been sorted by weight [19], a research team with no special interest in the Sun finally recognized that *atoms are sorted by mass in the Sun itself.*

Please send questions or comments to Prof. O. Manuel: om@umr.edu

3. http://web.umr.edu/~om/abstracts2005/Oxygen_to_Carbon_Ratio.pdf

4. http://web.umr.edu/~om/archive/SolarAbundances.pdf

5. http://web.umr.edu/~om/abstracts/gong-2002.pdf

6. http://web.umr.edu/~om/abstracts2003/jfe-superfluidity.pdf

7. http://www.berkeley.edu/news/media/releases/2004/06/01_maunder.shtml

8. http://epact2.gsfc.nasa.gov/don/00HiZ.pdf

9. http://www.copernicus.org/icrc/abstracts/ici6325.pdf

10. http://web.umr.edu/~om/archive/StrangeXenon.pdf

11. http://web.umr.edu/~om/abstracts2005/Noble_Gas_Anomalies.pdf

12. http://web.umr.edu/~om/archive/XenonRecord.pdf

13. http://web.umr.edu/~om/abstracts2001/windleranalysis.pdf

14. http://web.umr.edu/~om/abstracts2005/Nolte_and_Lietz.pdf

15. http://web.umr.edu/~om/abstracts2001/nuc_sym3.pdf

16. http://web.umr.edu/~om/abstracts2003/jfe-neutronrep.pdf

17. http://web.umr.edu/~om/abstracts2004/om-solar-neutrino.pdf

18. http://web.umr.edu/~om/abstracts2005/Iron_Rich_Sun.pdf

19. A pdf file of this 1980 review paper is in preparation.